\documentclass[10pt,conference]{IEEEtran}
\IEEEoverridecommandlockouts
\usepackage{cite}
\usepackage{amsmath,amssymb,amsfonts}
\usepackage{algorithmic}
\usepackage{graphicx}
\usepackage{textcomp}
\usepackage{xcolor}
\usepackage[utf8]{inputenc}
\usepackage{caption}
\usepackage{footnote}
\usepackage{hyperref}
\makesavenoteenv{tabular}
\makesavenoteenv{table}
\usepackage{listings}
\usepackage{framed}

\graphicspath{{figs/}}

\usepackage[backgroundcolor=blue!10,bordercolor=gray]{todonotes}

\def\BibTeX{{\rm B\kern-.05em{\sc i\kern-.025em b}\kern-.08em
    T\kern-.1667em\lower.7ex\hbox{E}\kern-.125emX}}
\begin{document}

\title{On the Rise and Fall of Simple Stupid Bugs: a Life-Cycle Analysis of SStuBs
\thanks{}
}

\author{
\IEEEauthorblockN{Balázs Mosolyg\'o\IEEEauthorrefmark{2}, Norbert V\'andor\IEEEauthorrefmark{2}, G\'abor Antal\IEEEauthorrefmark{2}, P\'eter Heged\H{u}s\IEEEauthorrefmark{1}\IEEEauthorrefmark{2}}
\IEEEauthorblockA{\IEEEauthorrefmark{2}\textit{Department of Software Engineering, University of Szeged},
Szeged, Hungary \\
\{mbalazs $|$ vandor $|$ antal \}@inf.u-szeged.hu}
\IEEEauthorblockA{\IEEEauthorrefmark{1}\textit{MTA-SZTE Research Group on Artificial Intelligence, ELKH},
Szeged, Hungary \\
hpeter@inf.u-szeged.hu}

\vspace{-20pt}
}

%\author{
%\vspace{49px}
%}

\maketitle

\begin{abstract}
Bug detection and prevention is one of the most important goals of software quality assurance.
Nowadays, many of the major problems faced by developers can be detected or even fixed fully or partially with automatic tools.
However, recent works explored that there exists a substantial amount of simple yet very annoying errors in code-bases, which are easy to fix, but hard to detect as they do not hinder the functionality of the given product in a major way.
Programmers introduce such errors accidentally, mostly due to inattention.

Using the ManySStuBs4J dataset, which contains many simple, stupid bugs, found in GitHub repositories written in the Java programming language, we investigated the history of such bugs.
We were interested in properties such as:
How long do such bugs stay unnoticed in code-bases?
Whether they are typically fixed by the same developer who introduced them?
Are they introduced with the addition of new code or caused more by careless modification of existing code?
We found that most of such stupid bugs lurk in the code for a long time before they get removed.
We noticed that the developer who made the mistake seems to find a solution faster, however less then half of SStuBs are fixed by the same person.
We also examined PMD's performance when to came to flagging lines containing SStuBs, and found that similarly to SpotBugs, it is insufficient when it comes to finding these types of errors.
Examining the life-cycle of such bugs allows us to better understand their nature and adjust our development processes and quality assurance methods to better support avoiding them.
\end{abstract}


\section{Introduction}
\label{sec:intro}

In our current climate of rapidly evolving expectations and needs towards software, fast development is a necessity.
This, however, comes at a cost: bugs are unavoidable.
Finding and fixing them have become an almost everyday activity to most developers.
This task is as important as it is difficult, and as such many bugs remain unnoticed for lengthy periods of time, with many bugs being noticed after they have already made their negative impact.
To prevent this in a critical software, in-depth testing must be performed to ensure correct functionality.
Despite the in-depth testing and code reviews, released software may still contain numerous bugs as the source code of such software systems are written by human beings.
A famous example of a simple, unnoticed bug having catastrophic consequences is the case of the ESA ARIANE 5 space shuttle launch of 11th December 2002, where a single wrong conversion caused millions of dollars worth of damage~\cite{lions1996ariane, nuseibeh1997ariane}.

Bugs like this are not as prevalent as others, making them harder to find.
A subset of these type of bugs are called simple stupid bugs~\cite{sstubs}, or SStuBs for short, which are minor inaccuracies that do not prevent the code from running or compiling, but change its behavior in a subtle way that may lead to unforeseen consequences.

To gain a deeper understanding about these issues we have analyzed the SStuBs found in a subset of the ManySStuBs4J~\cite{sstubs} dataset that contains over 25,000 instances of such bugs and their fixes, written in the Java programming language.
Out of these records we eventually processed 22,275, examining 10,168 individual commits.
We wanted to find out how and where SStuBs get introduced: are they simple mistakes the original author of the code overlooked, or were they caused by a different person?
Following this line of thought, we also wanted to know who fixes these issues typically, is it done by the same person who introduced them or someone else?
%Does it make any difference in fixing times whether or not the same person finds these issues or not?
Do the developers who made the mistakes find and fix them faster than others seeing that code possibly for the first time?

Karampatsis and Sutton~\cite{sstubs} examined whether SpotBugs\footnote{\url{https://spotbugs.github.io}} can detect SStuBs and found that SpotBugs flagged only 12\% of them.
However, it reported another 200 million possible bugs that makes this tool infeasible for detecting SStuBs.
We were curious if another static analyzer tool (namely PMD\footnote{\url{https://pmd.github.io}}) performs better in flagging SStuB lines, to have a more thorough understanding on how far or close tools already are to being able to detect these simple mistakes.
\section{Related Work}
\label{sec:related}

The study of software bugs and their life-cycle has quite a big literature, and has been an active research area for decades.
One of the first studies was conducted by Perry and Stieg~\cite{perry1993software}.
This research is as relevant as it was in 1993, as bug fixing activities are and always will be a significant part of every software engineer's work.

\begin{figure*}[htb!]
  \centering
  \captionsetup{justification=centering}
  \includegraphics[width=1.7\columnwidth]{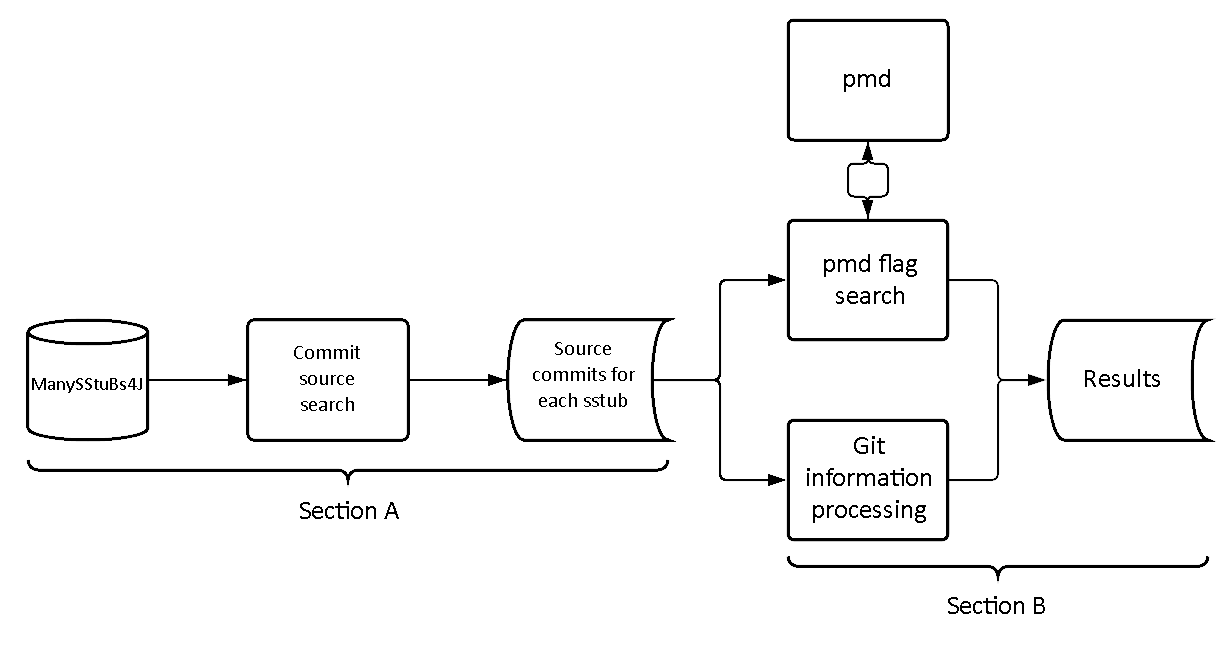}
  \caption{A visual representation of our process}
	\vspace{-10pt}
  \label{fig:process}
\end{figure*}

% Source code
Zimmermann et al.~\cite{4273265} mapped defects from the bug database of Eclipse to source code locations, they also added common complexity metrics.
Their other work~\cite{schroter2006if} maps failures to components, code, process, and developers.

% predict bug fixing time
Kim et al.~\cite{kim2006long} analyzed two open-source projects to compute bug fixing times.
They determined the fixing time by identifying bug-introducing changes (based on the work of Sliwerski et al.~\cite{sliwerski2005changes}).
%Their findings contain that bug fixing times range from 100-200 days.
Their results showed that bug fixing times range between 100 and 200 days.
Weiss et al.~\cite{weiss2007long} uses only previous bug report data to approach an unforeseen bug's fixing time.
Giger et al.~\cite{10.1145/1808920.1808933} also uses data from the issue tracker system to gain information about bugs and use the collected data to predict bug fixing times.

% Bug
Lamkanfi et al.~\cite{lamkanfi2010predicting} investigated whether the severity of a reported bug can be predicted accurately by analyzing its textual description.
They used open-source projects to validate their approach.

In his paper, Lucas D. Panjer~\cite{4228666} used Eclipse’s Bugzilla dataset to predict the lifetime of a bug (from the new to resolved bugs).
He used WEKA toolkit~\cite{hall2009weka} to perform data mining and analysis for prediction.

Saha et al.~\cite{6747164} analyzed long lived bugs in four open-source systems, and from five different perspectives: their proportion (between all bugs), severity, assignment (i.e where was most of the time spent in the bug fixing process), reasons (why these bugs were long lived) and the nature of fixes.

Other works analyze the life-cycle of vulnerabilities~\cite{frei2006large,wita2010ontology,joh2010framework,antal2020data}, a special type of bugs.

Although our work is similar to the above mentioned papers, namely we use code history to examine the life-cycle of bugs, we focus on and characterize a special subset of bugs, the SStuBs.
\section{Methodology}
\label{sec:methodology}

We analyzed bugs found in the ManySStuBs4J~\cite{sstubs} dataset, which are characterized by their simplicity and lack of obvious malfunction of the code.
We looked into their origin (i.e., the source commit introducing the SStuB) to analyze their life-cycle, and possibly gain a better understanding of them.
%Our motivating factor was the fact that the lack of understanding we face when it comes to such issues.
Figure~\ref{fig:process} shows an overview of our data analysis process.

\subsection{Finding the Source Commit}

The process of finding the origin of a SStuB\footnote{\url{https://github.com/MBalazs8796/MSR2021_LifeCycle}} was performed by the \texttt{git blame} command, which returns with the revision and author that last modified each line of a file.\footnote{\url{https://git-scm.com/docs/git-blame}}
%The main challenges we faced during the process of finding these source commits, was that some of the SStuB entries do not mark the correct line containing the error, and the fact that SStuB fixes can sometimes take considerable amounts of time, thus the source code in which they were found will most likely have changed by then.
We could not use the \texttt{bugLineNum} information directly to track the SStuBs as the exact position in the origin might be different than that in the code the SStuB was identified.
Therefore, we relied on the \texttt{bugNodeStartChar} and the \texttt{bugNodeLength} attributes in the entries of the dataset to find and store the SStuB-containing code blocks.
However, due to this it became ambiguous which of the lines in the block was the actual SStuB.
We handled this problem by looking through the file we know contains the SStuB, and searching for a matching code block (i.e., the code diff in the SStuB entry) in the parent files from the source commits.
This check is simple but time consuming, since it needs to be done for every potential source commit, of which there will be multiple in the cases where SStuB blocks appear.

\subsection{Collecting History Data}

The main challenge we faced while analyzing the history data was that to decide whether the SStuB was simply added in the source commit or it was introduced in the course of a modification of an existing line.
Our approach was to check the patch generated by the \texttt{git format-patch} command, and if the addition of the given line is preceded by a removal, it is treated as a modification.
We were conservative in calculating the statistics shown in this paper, meaning that any partial results except for the potential source commits' hashes have been omitted.

\subsection{Research Questions}

Our main goal with the examination of SStuBs was to find their sources and the way they are introduced and later fixed.
Aligned with this, we formulated four research questions, from which the first two is concerned with the introduction of SStuBs:

\begin{framed}
\vspace{-4pt}
\textbf{RQ1} Are SStuBs more likely to occur in code that is modified by multiple developers? 

\textbf{RQ2} Are SStuBs more likely to appear in newly added or modified code blocks?
\vspace{-5pt}
\end{framed}

%Knowing the average time elapsed between the introduction and eventual fix of a bug can help us better understand the importance of tools capable of finding them, if they are found within a couple of days of being introduced these tools are less necessary than if it took months to identify them.
Our third research question is connected with the removal, thus the end of life of a SStuB:

\begin{framed}
\vspace{-4pt}
\textbf{RQ3} How long does it take to fix a SStuBs, do authors notice their own mistakes faster?
\vspace{-5pt}
\end{framed}

Regarding the static analysis tools' capability of finding these issues, we thought that it would be interesting to know, how a static analyzer tool other than SpotBugs can protect against these kinds of bugs, leading us to our final research question:

\begin{framed}
\vspace{-4pt}
\textbf{RQ4} Can PMD flag SStuB lines as being error prone?
\vspace{-15pt}
\end{framed}
\section{Results}
\label{sec:results}

\begin{figure*}[htb!]
  \centering
  \captionsetup{justification=centering}
  \includegraphics[width=1.4\columnwidth]{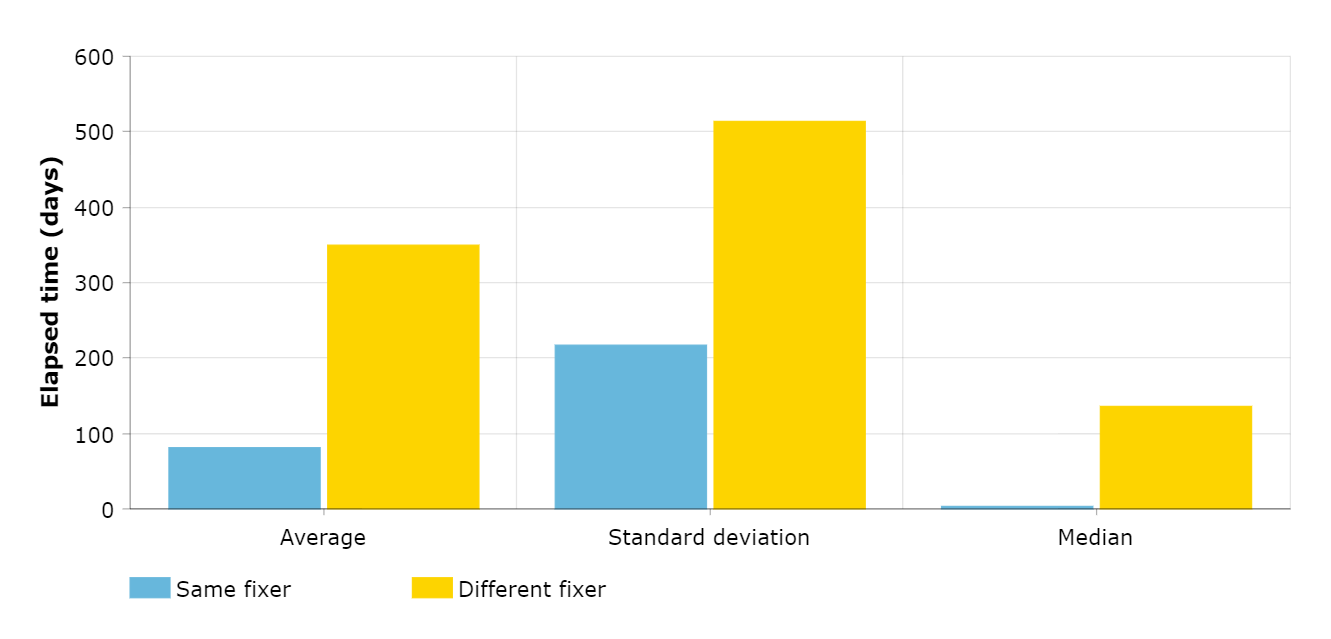}
  \vspace{-10pt}
	\caption{Figure demonstrating the difference in time elapsed between the creation and fix of SStuBs based on whether the same developer fixed it or not}
	\vspace{-15pt}
  \label{fig:time_stats}
\end{figure*}

When it comes to SStuBs, the main questions are the same as with any other bugs, how were they introduced, and how were they fixed.
These are important, since our ultimate goal is to either prevent or detect them fast enough, so that they do not cause any major issues.
%On our way towards these goals, it is useful to know some basic properties, such as whether more frequently changed code is more prone to SStuBs or not, or does it matter if only one developer wrote a part of a code, so we can decide what factors contribute to certain types of issues.

%It is also important to get a measure of how current solutions fair up against the problems we are looking to fix, since they might already be adequate at the task.

\subsection{The Nature of SStuB Introduction}

In predicting potential issues such as SStuBs, or even major bugs, it is a reasonable idea to put an emphasis on code that is changed frequently, and/or is modified by multiple developers since discrepancies in their goals or even just coding styles could lead to a higher potential for bugs.
We examined these ideas, first by checking whether most SStuBs were really written by a different developer than the surrounding lines, secondly by checking if the SStuB was introduced in the same commit as its surroundings (i.e., introduced in a newly added code block or by modifying an existing, previously correct code part).
We looked at buggy code blocks marked by the start character and length of a SStuB in the dataset.

We found that in 90.35\% of the cases the developer who created the surrounding lines also introduced the SStuB, and that in 76.22\% of cases the SStuB was created in the same commit as the surrounding lines.
These figures show that when it comes to the SStuBs presence in the dataset, in the vast majority of the cases the bugs were created by the same developer who worked on the neighboring lines as well, meaning that these minor issues were likely caused by a slight oversight on the developers part, and not by some other external factor.

\begin{framed}
\vspace{-4pt}
\textbf{RQ1} The SStuBs typically do not occur in code modified by multiple developers, in fact they seem to appear more in larger chunks of code written by the same developer, meaning that these kinds of mistakes are more likely to be due to the loss of attention rather than misunderstanding the code functionality.
\vspace{-5pt}
\end{framed}

Based on the observed majority of SStuBs that were written in the same commit as their surrounding counterparts, we can answer another research question.

\begin{framed}
\vspace{-4pt}
\textbf{RQ2} Most SStuBs are added in the same commit as their neighboring lines, meaning that they are not added at a later stage of development rather, when the block is first added to the code-base.
Since these bugs are rarely introduced in later modifications, it is safe to assume that frequent changes do not increase the likelihood of SStuBs significantly.
\vspace{-5pt}
\end{framed}

%Given that we looked at the strictest definition of surroundings possible, we can say that the difference in a line's source commit is a better deciding factor than the developer who created it.

\subsection{A closer look at SStuBs' lifetimes}

A good measure of how hard it is to find SStuBs is the time it takes to fix them.
Given that they are "simple stupid bugs", it is safe to assume that they are not difficult to fix, and as such are patched as soon as they are found.
In the case of bugs present in the dataset, the average time taken to fix a SStuB is approximately 240 days, however, the median time is only 58 days, with a standard deviation of 440 days.
It is clear that some of them are significantly harder to find, thus hide in the code-base much longer than the average.
However, even the relatively small median is large given the simplicity of these bugs, and leads to more questions.
One of those questions is, do developers not check their code after it is written?
The answer is, of course they do.
When the fix times are grouped based on if the same developer fixed the problem who introduced the SStuB or not, it becomes clear that when developers double-check their own code, they find these issues quicker than if someone was to look for a bug.
As can be seen in Figure~\ref{fig:time_stats}, when the same developer fixes the SStuB they wrote, they do it on average 3 times faster, in only 81 days, and with an even lower median of only 4 days.
In the case when someone else handles the problem, the average increases to 349 days, and the median also jumps to 136 days, showing that developers find issues in their own code faster than others.

The significant difference in these statistics tells us that it would be favorable for developers to look for these issues themselves, however, data shows that this is not a usual case.
Of the SStuBs found in the dataset, only 40\% were fixed by the same author who introduced them, leading to the overall increase in the time it takes to fix SStuBs seen above.

\begin{framed}
\vspace{-4pt}
\textbf{RQ3} On average, SStuB fixes take around 240 days, which is too long for any bug to stay unnoticed.
Around 40\% of them seem to be relatively quickly noticed by the developer who introduced them, hinting at the possibility that double-checking ones own code is the fastest way to get rid of the potential SStuBs.
\vspace{-5pt}
\end{framed}

\vspace{-8pt}
\subsection{Can a Static Analyzer Find SStuBs?}
We were curious whether only the SpotBugs tool was unable to efficiently find SStuBs, or it is simply too difficult task to a static analyzer.
To answer this question, we used PMD, a static analyzer widely used in Java projects.
In the end, not a single SStuB was flagged by PMD, further reinforcing the idea that these bugs are particularly hard to find and mostly undetectable by static analyzers.

\begin{framed}
\vspace{-4pt}
\textbf{RQ4} Lines with SStuBs are not considered error prone by neither SpotBugs~\cite{sstubs}, nor PMD, while these tools are not representative of every static analyzer, we can at least say that these kinds of bugs are difficult to find by static analyzers.
\vspace{-5pt}
\end{framed}

\section{Threats to Validity}
\label{sec:threats}

We were limited by the resources available, and so we could only work on the smaller version of the dataset, processing only 25,539 SStuB records instead of the total available 153,652. 
This impacts the accuracy of our research, but not the overall conclusion.
To confirm this, we looked through a smaller subset of our results, and the overall values have not changed significantly, leading us to the conclusion that the patterns we observed are consistent.

We used a strict interpretation of neighboring lines, when we created the statistics that involve surroundings, namely the \texttt{bugNodeLength}, which does not always contain extra lines.
We chose this approach instead of selecting a fixed surrounding size ourselves because this way we stay as close to the representation in the dataset as possible.
If we were to chose a fixed surrounding size, we may have examined parts of blocks completely irrelevant to the SStuB at hand.
\section{Conclusion}
\label{sec:conclusion}

We have looked through the life-cycle of bugs in the ManySStuBs4J dataset in order to gain a better understanding of their behavior when it comes to their introduction and the time it takes to fix them.
We found that most SStuBs are introduced by the same developer who created the surrounding nodes, meaning that most of the times they are not due to a lack of understanding the context in which the issue was created, rather a loss of attention while adding new code.
A majority of SStuBs are created in the same commit as their surroundings, pointing us again towards the conclusion described above.
However, the likelihood of the SStuBs being introduced in a different commit is higher than it being made by a different developer, meaning that some SStuBs are caused by later modifying potentially correct code.

We also found, that most SStuBs take a significant time to find and fix, which reinforces the idea of them being hard to find, while barely having a noticeable impact on functionality.
We showed that the fix times of SStuBs are significantly shorter in the cases when the same developer who introduced also fixes them.
Unfortunately, the majority of the SStuBs are not fixed by the same developer who introduced them.

We also looked at whether another static analyzer would mark any of the lines containing SStuBs as error prone or not, but found that PMD was unable to mark any SStuBs in the analyzed dataset.

\vspace{-10pt}
\section*{Acknowledgment}
The presented work was carried out within the SETIT Project (2018-1.2.1-NKP-2018-00004)\footnote{Project no. 2018-1.2.1-NKP-2018-00004 has been implemented with the support provided from the National Research, Development and Innovation Fund of Hungary, financed under the 2018-1.2.1-NKP funding scheme.} and supported by the Ministry of Innovation and Technology NRDI Office within the framework of the Artificial Intelligence National Laboratory Program (MILAB).

Furthermore, Péter Hegedűs was supported by the Bolyai János Scholarship of the Hungarian Academy of Sciences and the ÚNKP-20-5-SZTE-650 New National Excellence Program of the Ministry for Innovation and Technology.

\bibliographystyle{./IEEEtran}
\bibliography{bibl}

\end{document}